# Surface phases of the transition-metal dichalcogenide IrTe$_2$

Chen Chen, Jisun Kim, Yifan Yang, Guixin Cao, Rongying Jin, and E. W. Plummer[1]*

[1]Department of Physics & Astronomy, Louisiana State University, Baton Rouge, Louisiana 70803, USA

**Abstract**

Transition-metal dichalcogenide IrTe$_2$ has attracted attention because of striped lattice, charge ordering and superconductivity. We have investigated the surface structure of IrTe$_2$, using low energy electron diffraction (LEED) and scanning tunneling microscopy (STM). A complex striped lattice modulations as a function of temperature is observed, which shows hysteresis between cooling and warming. While the bulk 5×1 and 8×1 phases appear at high temperatures, the surface ground state has the 6×1 phase, not seen in the bulk, and the surface transition temperatures are distinct from the bulk. The broken symmetry at the surface creates a quite different phase diagram, with the coexistence of several periodicities resembling devil's staircase behavior.

**Introduction**

Transition-metal dichalcogenides (TMDs) with general formula MX$_2$ (M = transition metal, X = chalcogen element) form a layered structure, which have been extensively studied because of their intriguing properties, such as the coexistence or competition of charge density wave [1-5] and superconductivity [6-11]. The inherent layered crystal structure has also made it possible to study monolayers and bilayers of these compounds [12-15]. While there are many potential applications, the fundamental properties of these TMDs are yet to be fully understood, particularly in the circumstance of spatial confinement. For example, many of phase transitions are accompanied with the formation of one-dimensional (1D) striped ordering, which breaks the inherent high-temperature symmetry [4,16-19]. What is the response of surface which already breaks the translational symmetry?

IrTe$_2$ is such a compound, whose bulk crystal structure at 300 K is shown in Fig. 1(a) with its top view in Fig. 1(b). The triangular Ir layer is sandwiched between two Te layers with the stacking sequence "*1T*", where "*1*" stands for the number of Te-Ir-Te triple layers in a unit cell and "*T*" stands for the trigonal unit cell symmetry [20]. At room temperature the unit cell can be defined by the three vectors *a*, *b*, and *c* in Fig. 1(a), which called the 1×1×1 phase. Cooling produces multiple phase transitions associated with the coupling between the charge and the lattice [6,9,21,22]. In bulk, there is a first-order structural transition from 1×1×1 phase into a 5×1×5 phase at ~ 280 K, a structure with five times the periodicity both in the *a* and *c* directions. The existence of long-range ordering along the *c* direction means that there is significant interplanar coupling, i.e., the bonding between planes is not simple van der Waals type. It has been proposed that the structural transition originates from the Ir-Ir dimerization in the *ab*-plane, with the Ir atoms moving towards each other, lowering the energy [23,24]. The dimers form striped ordering in the *ab*-plane as displayed in Fig. 1(c). The stacking sequence of the striped phase has the same periodicity along both the *c*-axis direction and *a*-axis directions (5×1×5). As stated previously, this is unexpected for a weakly bounded layered material [23-25]. Such ordering turns the high-temperature trigonal phase into the low-temperature triclinic phase, with new unit cell vectors orientated in entirely different directions (the new *c*-axis is not perpendicular to *a* and *b*) [6,26-28]. Further cooling causes the second structural transition at ~ 180 K, forming an 8×1×8 structural modulation with additional dimers in the unit cell [21,29]. Warming the 8×1×8 structure produces only one structural transition at ~ 280 K, where the 8×1×8 structure converts directly to the 1×1×1 structure [21]. It is relevant to our observations that when Te is partially substituted by Se, a 6×1×6 modulation can be stabilized at room temperature, with the monoclinic unit cell and a dimerized structure [29,30].

Creating a surface breaks the translational symmetry and disturbs the coupling along the *c*-axis direction, so new phases are expected to emerge. It is difficult, using surface techniques, to observe the *c*-axis stacking sequence. Therefore, with the bulk 5×1×5 phase, all that can be seen is a 5×1 surface phase. The surface properties of IrTe$_2$ have been investigated using STM, which probes the local charge density but can infer the structural symmetry [17,18,31-33]. In contrast to the bulk, the fully dimerized 6×1 modulation [Fig.

1(d)] can be stabilized at low temperatures at the IrTe$_2$ surface [16,18]. Hsu *et al.* [17] reported 11×1 and 17×1 surface modulations at low temperatures not seen in the bulk, with the tendency to generate a (3*n*+2) structure, where *n* is an integer indicating number of dimers in a unit cell. It is obvious that the surface has different symmetries than the bulk as seen by STM. However, STM does not directly probe the atomic structure. To date, the only surface structural measurement is via low energy electron diffraction (LEED), which showed a 5×1 reconstruction at 200 K and 1×1 character at 300 K [28]. Here, we report the temperature dependence of IrTe$_2$ surface structures studied by both LEED and STM. During the cooling process, there are structural transitions from 1×1 to 5×1 phase at ~ 280 K, then to intermediate phase dominated by 8×1 at ~ 170 K and to predominately 6×1 phase below 85 K. During the warming process, the low-temperature 6×1 phase converts mainly to 8×1 at T > ~ 200 K with the appearance of 5×1 structure near 288 K, and finally back to 1×1 at 300 K. Both LEED and STM observe coexistence of multiple striped orderings, resembling the "devil's staircase" behavior.

A simple model describing the formation of 1D atomic chains under external potential is the Frenkel-Kontorova model, whose solution includes the coexistence of different phases separated by domain walls (solitons) [34]. When phases with different periodicities coexist without a dominant single phase because of their nearly degenerate energies the behavior can be associated with the "devil's staircase", which resembles the low-temperature surface features. The competing phases can be tuned by external parameters such as temperature, pressure and magnetic field [35-38].

**Experimental Techniques**

Single crystals used in these experiments were grown by the self-flux method, with the procedure described previously [9] [39]. All samples were cleaved at 300 K in ultra-high vacuum (UHV) environment with base pressure < 1×10$^{-10}$ Torr, and subsequently cooled to 85 K with liquid nitrogen or 45 K with liquid helium. All the warming processes started with the sample at 85 K. LEED measurements were taken immediately after cleaving. The STM images were acquired at 300 K after cleaving and 80 K after cooling.

**Results**

Samples cleave between the Te-Te layers due to its layered nature as shown in Fig. 1(a). The two Te layers are equivalent in the bulk, but the broken symmetry at the surface can make the Te1 layer different from the Te2 layer. The cleaved surface with bulk truncation is shown in Fig. 1(b), where the surface unit cell is indicated by the black rhombus and labeled 1×1. The 2D space group of the surface structure is $p3m1$.

When the sample is cooled below 280 K, the structure becomes striped presumably as a result of the Ir-Ir dimerization [23]. If two dimerized rows are separated by a single undimerized row of Ir atoms, the dimerized segments form a 3× periodicity. If the dimerized segments are separated by two undimerized rows, it forms the (3$n$+2) structure. Fig. 1(c) shows the top view of the 5×1 striped surface corresponding to $n = 1$, with the red shade indicating the Ir-Ir dimers. The 5×1 unit cell is marked by the black parallelogram in Fig. 1(c) with the 2D space group $p1$. At the lowest temperature, the structure stabilizes into a 6×1 structure, which is consistent with dimerized segments occupying the whole surface as shown in Fig. 1(d) [18]. Here the adjacent two 3× dimerized rows have dimers rotated 120° with respect to each other, the unit cell is 6×1 contains two dimers, presented by the black parallelogram in Fig. 1(d). Given this structure, there is a glideline symmetry indicated by the blue dashed line in Fig. 1(d), producing a 2D space group $pg$.

To aid in the interpretation of the LEED data, simulated LEED patterns (based on symmetry) for different striped surface structures are shown in Figs. 2(a)-(e). Figure 2(a) shows a 1×1 pattern for the undistorted structure in Fig. 1(b). The reciprocal unit vectors are shown as a* and b*. The spot at the center is defined (0,0), the spot at a* is (1,0), and the spot at b* is (0,1). All the spots on a 1×1 pattern can be labeled by integer numbers using a* and b*. Figure 2(b) shows the LEED pattern for a 5×1 structure, where there are four additional fractional spots between (0,0) and (1,0) spots. The parallelogram constructed by the red and green lines in Fig. 2(b) represents the reciprocal unit cell from a 5×1 real space unit cell in Fig. 1(c). Due to the dimmer formation, there exists three domains at the surface, which are 120° with respect to each other [32], and each domain there forms 1D striped ordering. The corresponding LEED pattern is a superposition of patterns from the three domains, shown in Fig. 2(c). The simulated 8×1 and 6×1 patterns are shown in Figs. 2(d) and (e). There should be no glideline in the 6×1 surface structure, which will lead to missing spots in a LEED pattern.

Figures 2(f)-(o) show the LEED patterns of the $IrTe_2$ surface at indicated temperatures, all taken at a beam energy of 80 eV, which maximizes the intensity of the fractional ordered spots. The freshly cleaved $IrTe_2$ surface shows a well-defined 1×1 pattern at 300 K [Fig. 2(f)], consistent with Fig. 2(a). The three spots of one equilateral triangle have the same intensity but different from the other triangle, i.e. the trigonal symmetry. This feature is better resolved at 122 eV, as shown in Fig. 2(p).

Upon cooling, fractional spots appear as seen at 271 K in Fig. 2(g) and 213 K in Fig. 2(h). This 5×1 reconstruction pattern is consistent with the simulated LEED pattern with the presence of three domains in Fig. 2(c), and with previous LEED results [28].

Further cooling to 160 K produces 8×1 reconstruction as shown in Fig. 2(i). Reducing the temperature to 85 K, the surface undergoes a transition to the 6×1 structure, shown in Figs. 2(j) and 2(k) which are consistent with that shown in Fig. 2(e). This pattern is also consistent with the Fourier transform of the $IrTe_2$ topographic STM image at 4.5 K [18]. The glideline symmetry (Fig. 1(d)) induces structural factor cancellation, causing the fractional spots with odd order to be missing along the ($a^*$,0) direction. These missing spots are indicated by red color in the simulated LEED pattern of Fig. 2(e). This glideline symmetry seen in the data [Figs. 2(j)-(k)] confirms the dimer arrangement with different orientation shown in Fig. 1(e). If dimers are oriented in the same direction, there is no glideline symmetry and the reconstructed unit cell should be 3×1. The spots indicated by yellow color in Fig. 2(e) are also missing in the 6×1 LEED pattern in Figs. 2(j) and 2(k). This is because their intensities are small at 80 eV, but they can be clearly resolved at 66 eV and 87 eV (Appendix A Fig. 7). Our LEED result of the 6×1 lattice structure indicates a higher transition temperature than previously reported using STM [16-18].

When the sample is warmed, there is a large hysteresis in the transitions at the surface, just as in the bulk. Figures 2(l)-(o) are collected during the warming process from 85 K. At 160 K and 213 K the LEED patterns show 6×1 feature, unchanged from low temperatures. Continued warming changes the surface structure into a 8×1 reconstruction at 271 K [Fig.

2(n)]. Finally, the structure converts to the original 1×1 phase after warming up to 300 K (Fig. 2(o)).

The 6×1 and 8×1 phases coexist over a wide temperature range as shown in Fig. 3. To better resolve these phases, we apply the electron beam is off the normal. Figure 3(a) is such a LEED pattern at the same temperature and beam energy as in Fig. 2(i), There are three clearly resolved spots between (0,0) and (1,0) spots, and the line profile analysis based on red line in Fig. 3(a) shows these three spots are (2/8,0), (3/8,0), and (5/8), in agreement with an 8×1 structure (Appendix B Fig. 8). The LEED pattern in Fig. 3(b) is collected after laterally shifting the sample position ~ 1 mm from Fig. 3(a). It exhibits 6×1 features between (0,0) and (1,0) spots. Both phases can also coexist at the same part of the surface, even at the lowest measured temperature, as demonstrated in Fig. 3(c). The beam energy here is 55 eV and the main feature along the direction indicated by the arrow at the lower right corner is 6×1. The 8×1 feature exists along the direction indicated by the arrow at the lower left corner of Figs. 3(c) and (d). The implication is that the 6×1 feature is most likely the ground state of the surface, but part of the sample surface still has the remaining 8×1 structure due to the bulk. During the warming process, similar coexistence is also observed (Fig. 3(d)), which is taken at a different sample position than that of Fig. 2(m). The bulk measurements on the sample used in the LEED experiments show that there is no coexistence of multiple phases at any temperature.

Figure 4(a) shows the STM topographic image of a freshly cleaved IrTe$_2$ surface at 300 K, which has a hexagonal symmetry associated with the top layer Te atoms. When the sample is cooled down to 80 K, the striped ordered surface is shown in Fig. 4(b). This large area STM image shows the coexistence of two different features in two domains: mixed striped area (upper right) and uniform striped area (lower left). In the mixed striped area, there is no predominant periodicity. This is illustrated by the atomically resolved image in Fig. 4(c), where the striped rows with different contrast coexist with no repeating order. An atomically resolved image from the uniformly striped area is shown in Fig. 4(d), displaying the striped structure with dimers that have a 6×1 structure, consistent with LEED results and Ref. [18]. It also confirms that the known ground state 6×1 phase remains at 80 K, a temperature higher than previous STM observations.

In the bulk there is only one phase transition upon warming, compared to two for cooling [9,17,21,30]. The LEED pattern at 288 K during warming process shows 5×1 feature in Fig. 5(a), although the spot intensity is smaller compared to the cooling process. The sample surface undergoes a transition from 8×1 to 5×1, and finally to 1×1 in a small temperature window. Figure 5(b) is a magnified image of the red square in Fig. 5(a) including three fractional spots [(2/5,0), (3/5,0), and (4/5,0)] and one integer spot [(1,0)]. The three fractional spots are elliptical in shape while the integer spot is circular. The analysis of the linewidths of the diffraction spots is displayed in Fig. 5(c), showing the line profiles for the five cuts in Fig. 5(b). All the line profiles are normalized based on the pixels in the image, and the peaks are fitted with Lorentzian functions. The fitted results are summarized in Table 1. The linewidths from peaks A, B, and C of cut 5 are almost twice as large compared to the five other peaks. The broadening of the diffraction spot can be understood in a way similar to the Williamson-Hall plot in X-ray scattering, where the linewidth of the diffraction peak is inversely proportional to the mean crystallite size [40]. For the $IrTe_2$ surface, the lack of long range ordering of the 5×1 feature during the warming process leads to the increase of fractional spots' linewidths. The size of the 5×1 domain can be estimated to be smaller than the probing electron's coherence length, which is around 100 Å [41]. This may explain why the 5×1 structure is not seen during warming in the bulk [21].

**Discussion**

The observed surface phases are summarized in Fig. 6, and compared to the bulk phases. At room temperature, the surface shows a 1×1 pattern. With slight cooling, the 5×1 pattern emerges at a temperature around ~ 280 K, consistent with the bulk structural transition. With further cooling, the 5×1 pattern disappears at temperatures lower than ~ 170 K, and the surface transforms into an 8×1 phase. The 6×1 phase gradually emerges together with the 8×1 phase upon further cooling. As it persists even at the lowest measured temperature (45 K), we propose that the surface ground state has the 6×1 structure (Fig. 6(a)). The warming process shows different characters, which are summarized in Fig. 6(b). The 6×1 ground state persists up to the temperature higher than 200 K, then transforms to the 8×1 phase at ~ 200 K and persists to almost ~ 280 K. The unique signature at the surface is the reemergence of the ordered 5×1 phase observed near 288 K, which is not observed in the bulk during the warming process [21].

The rich phases of IrTe$_2$ emerge as a result of the competition between various degrees of freedom such as charge and lattice. The creation of surface alters the balance, thus generating new phases. For example, the 6×1 surface state can be similarly achieved in the bulk by substituting the Te atoms with Se [29]. One possible reason is the difference in coupling strength between the triple layers along the *c*-axis direction. In the bulk the inter-layer bonding causes lattice distortion along the out-of-plane direction when the dimers are formed at low temperature [23,29]. The surface breaks the bulk translational symmetry, resulting in a unique environment for the top "triple layer" with different coupling along the *c*-axis.

The coexistence of numerous spatially modulated phases resembles the devil's staircase behavior at the surface, which might be the intrinsic property of the 2D layer. These phases must have similar energies, with competition between two periodicities [37,42]. Ideally the devil's staircase behavior would be observed through the coexistence of phases along the same direction. However, due to the limited LEED resolution, it is difficult to distinguish higher ordered (3$n$+2) patterns when $n > 2$, even more difficult when multiple ordering coexists along the same direction. STM resolves this problem by measuring short-range ordering displayed in Fig. 4(b), confirming the coexistence of various ordering in local area at the surface.

**Conclusion**

In summary, we have used LEED and STM to investigate the temperature dependent surface phases of IrTe$_2$. The surface stabilizes fully dimerized 6×1 phase at low temperatures, distinctly different from the bulk ground state. The cooling and warming processes exhibit hysteresis in the surface phase diagram with richer features than the bulk. Our results imply the surface phases are consistent with a devil's staircase behavior. The fact that the surface behavior of this TMD is different than the bulk has significant impact on the use of 2D sheets of TMDs for device application [43].

**Acknowledgement**

Primary support for this project came from the National Science Foundation Grant DMR-1504226, with crystal growth effort supported by the U.S. Department of Energy under EPSCoR grant DE-SC0012432 and Louisiana Board of Regents (Guixin Cao).

## Appendix A: Analysis of 6×1 LEED pattern

These missing spots due to the glideline symmetry in 6×1 LEED pattern are indicated by red colors in Fig. 2(e). They are always missing because the vanishing intensity is from the glideline symmetry. In contrast, the spots indicated by yellow color in Fig. 2(e) are not truly missing, instead their intensities are small at 80 eV. These spots can be clearly resolved at 66 eV and 87 eV. These spots are labeled (0.5,1) and equivalent, shown in Fig. 7.

## Appendix B: Analysis of 8×1 LEED pattern

Detailed analysis of the 8×1 pattern is explained in Fig. 8. Figure 8(a) is a reproduce of Fig. 3(a), with the line profile shown in Fig. 8(b) after subtracting a linear baseline. There are five predominant peaks which are fitted by Lorentzian functions. The $x$-axis of the line profile is normalized by setting the first and last peak positions to be 0 and 1, respectively. By adapting this line profile into an expected 8×1 pattern, the three peaks in between correspond to (2/8,0), (3/8,0), and (5/8) spots. The positions of all five peaks and their peak numbers are plotted in Fig. 8(c) as the black balls. These five spots can be fitted perfectly by a straight line in Fig. 8(c), with red balls locate at the correct positions of the missing peaks as guide to the eye. Fig. 7(d) is the LEED pattern at the same temperature and beam energy as Fig. 2(n) but the sample is rotated. It has the same features of a 8×1 pattern such as the peak positions and missing of several spots.

The missing spots (1/8,0), (4/8,0), (6/8,0), and (7/8,0) are not from the structural factor cancellation. Their beam intensities are small at this energy due to the diffraction condition similar to the yellow spots in Fig. 2(e). For instance in Fig. 8(d) the (1/8,0) spot in one domain seems missing but in another domain is evident. Compared to other types of patterns the 8×1 surface is worse in sharpness, implying that the 8×1 phase does not have long range ordering as other phases. Within an energy range of 40 to 400 eV there are no spots that never emerge, but the missing ones at 80 eV always have smaller intensities than the existing ones. This could offer some insight for the structural determination of the 8×1 phase through the structural factor calculation. It can be speculated based on the asymmetric diffraction pattern that the new unit cell after 8×1 reconstruction has broken inversion symmetry compared to undistorted unit cell.

# Tables

Table 1: Linewidths from the Lorentzian fittings in Fig. 5(c). The linewidths of the peak A-C in cut 5 are much larger than peak D and cut 1-4.

|  | Cut 1 | Cut 2 | Cut 3 | Cut 4 |
|---|---|---|---|---|
| Linewidth | 15.0±0.3 | 12.1±0.3 | 16.3±0.7 | 15.7±0.5 |
| Cut 5 | A | B | C | D |
| Linewidth | 33.6±1.2 | 30.1±1.5 | 26.6±1.3 | 13.1±0.4 |

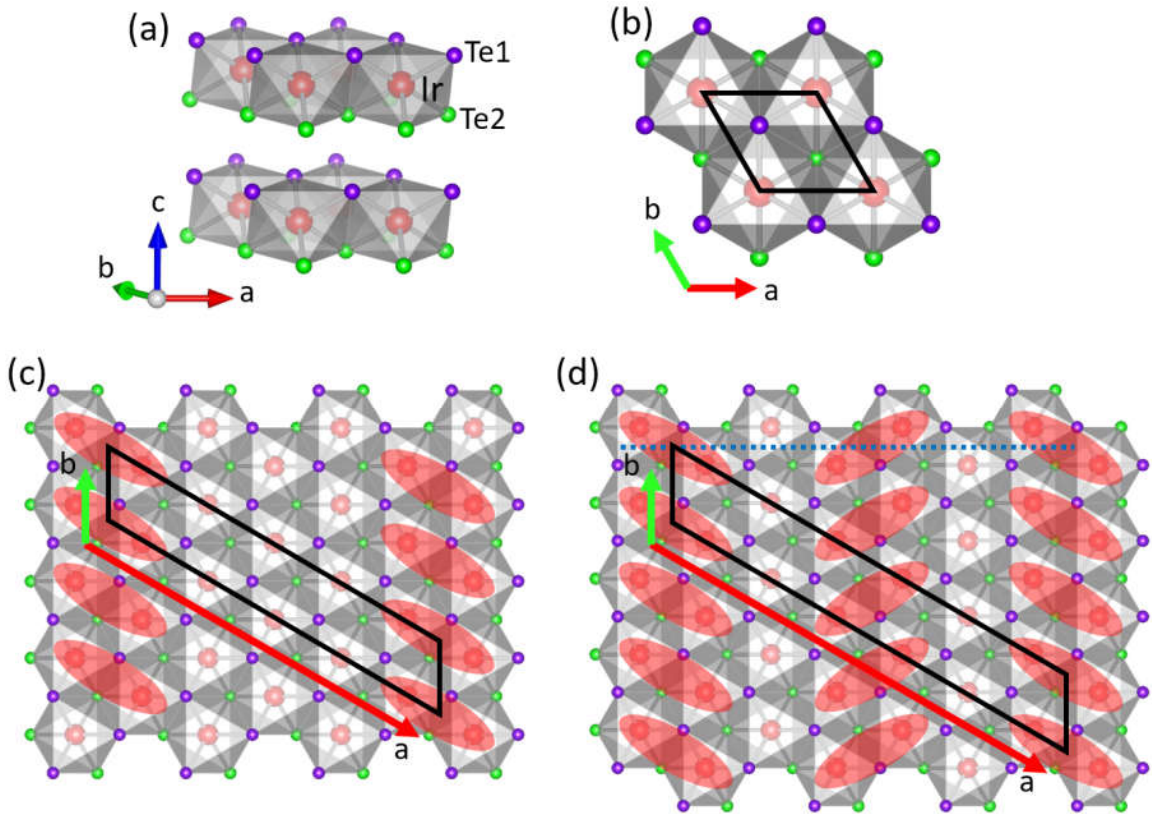

FIG. 1 (color online). Crystal structure of IrTe$_2$: (a) bulk structure, where two Te atoms labeled in purple and green are equivalent; (b) Bulk truncated surface with the undistorted 1×1 unit cell; (c) Reconstructed surface lattice with the 5×1 structure with Ir-Ir dimers shown by red shades; (d) Surface 6×1 structure with the blue dashed line indicating the glideline symmetry.

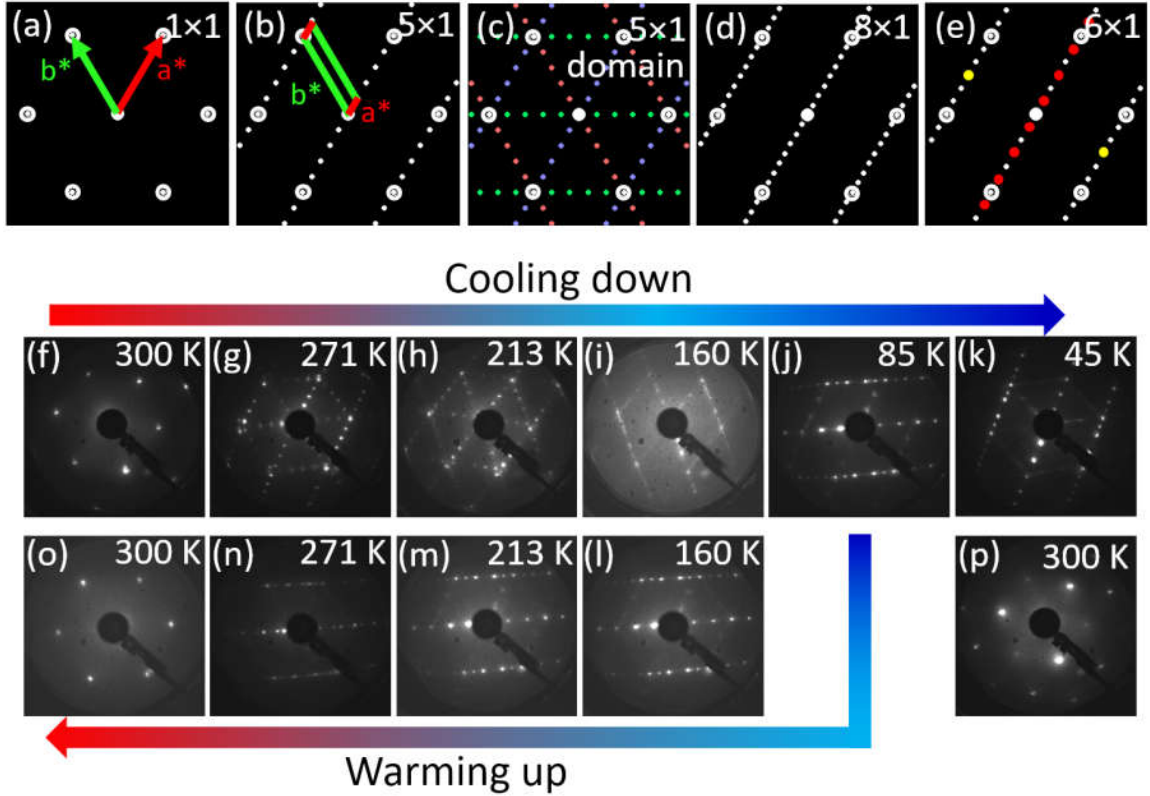

FIG. 2 (color online). (a)-(e) Simulated LEED patterns for different structures with integer spots in white circles: (a) undistorted 1×1 structure, (b) 5×1 structure with single domain, (c) 5×1 structure with multiple domains, (d) 8×1 structure, and (e) 6×1 structure. (f)-(o) LEED patterns at indicated temperatures during the cooling and warming processes. The patterns are taken at 80 eV; (p) 1×1 structure at the same temperature as (f), but with 122 eV beam energy.

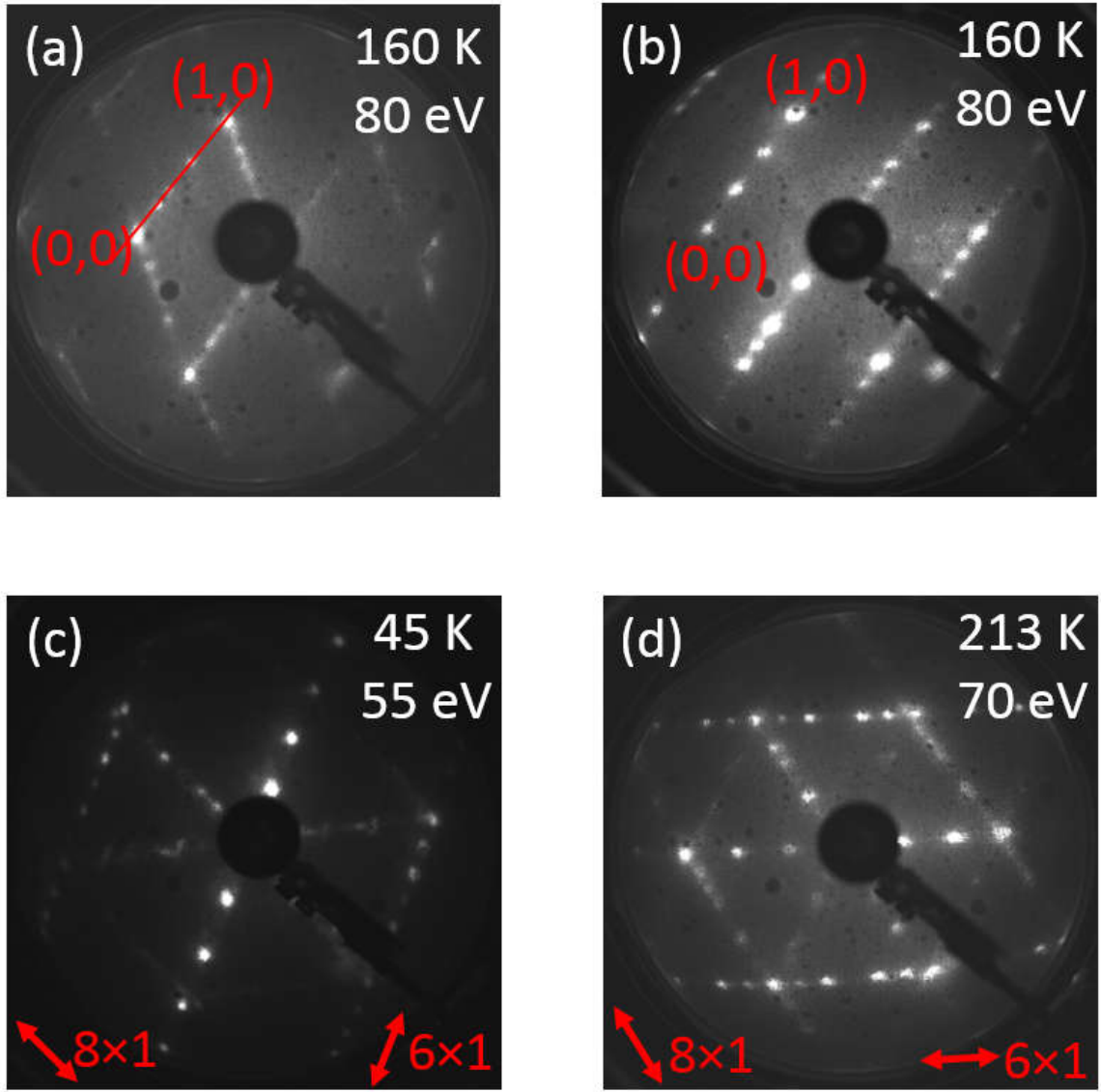

FIG. 3 (color online). (a) LEED pattern at 160 K and 80 eV taken after a rotation and lateral translation of ~ 1 mm from Fig. 2(i); (b) LEED pattern at 45 K and 55 eV taken at the same position as Fig. 2(k); (c) LEED pattern at 213 K and 70 eV with a slight different position from Fig. 2(m). The corresponding domains are marked by the arrows at the bottoms of (c) and (d).

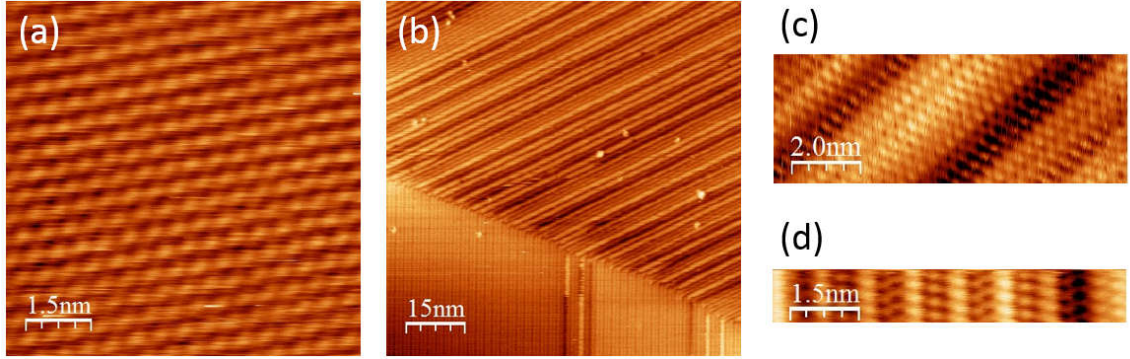

FIG. 4 (color online). Topographic STM images of IrTe$_2$ surface. (a) Atomically resolved image of a freshly cleaved surface at 300 K, scanned at V$_{sample}$ = 1 V, I = 80 pA. The surface shows uniform 1×1 hexagonal structure. (b) Large area image at 80 K, scanned at V$_{sample}$ = 2.0V, I = 80 pA. It shows the coexistence of mixed striped area and uniform striped area; (c) Atomically resolved image representative of the mixed striped area, scanned at V$_{sample}$ = -0.6 V, I = 50 pA; (d) Atomically resolved image representative of the uniform striped area with 6×1 reconstruction, scanned at V$_{sample}$ = -5 mV, I = 5 nA.

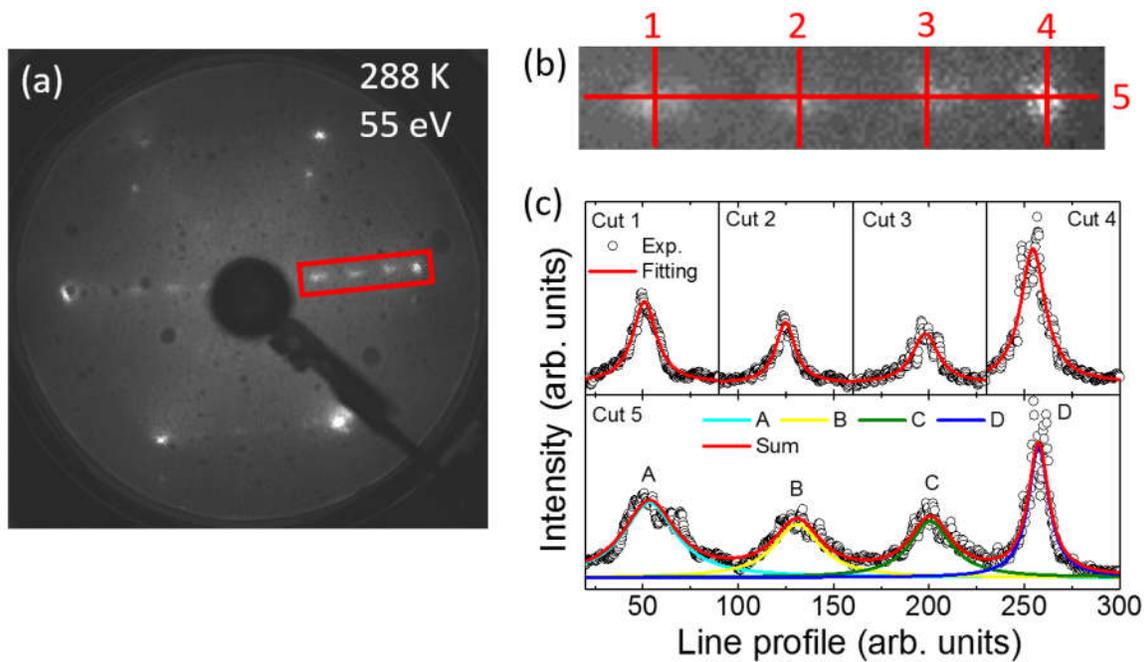

FIG. 5 (color online). (a) LEED pattern at 288 K 55 eV during the warming process; (b) Zooming in the area indicated by the red square in (a); (c) Line profile data from the cuts in (b); The coordinates of $x$ axis are normalized for all data based on the image pixels.

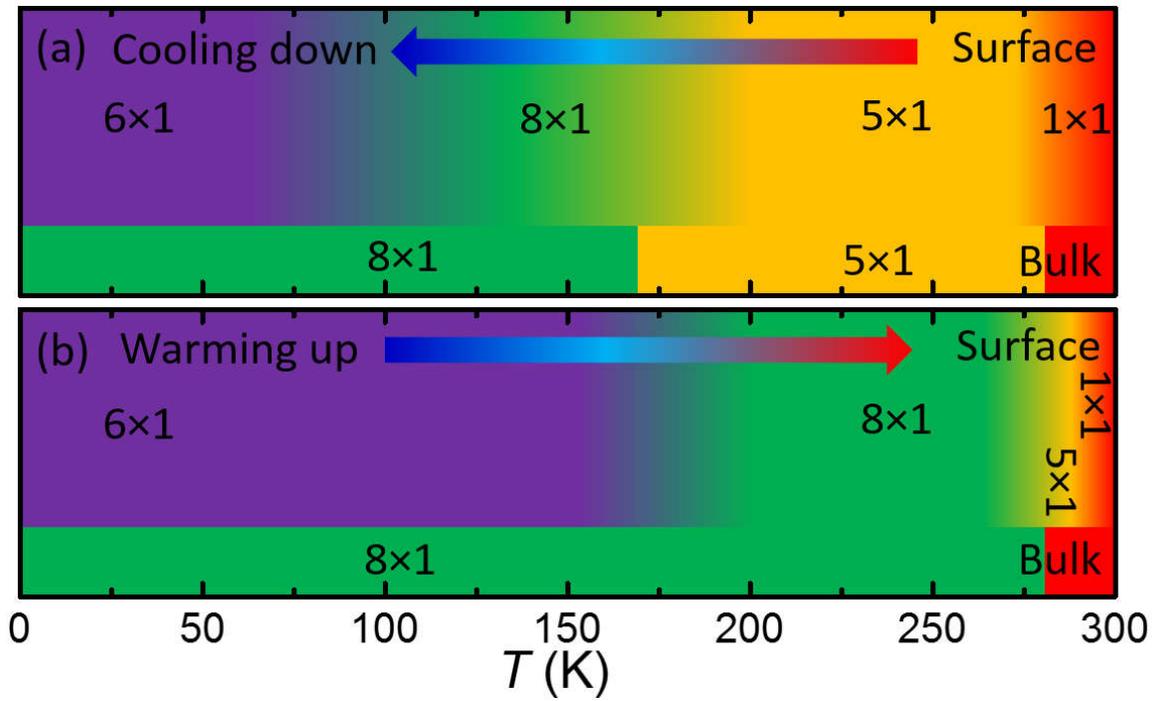

FIG. 6 (color online). Comparison of surface and bulk structures at different temperatures for (a) cooling process and (b) warming process.

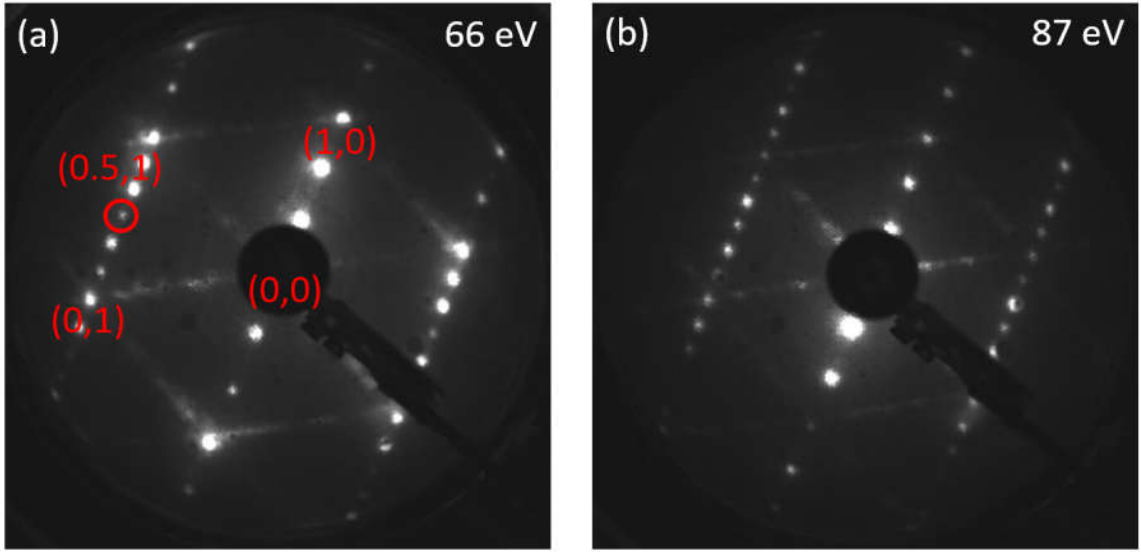

FIG. 7 (color online). LEED patterns of IrTe$_2$ at 45 K and different electron beam energies. These patterns show emergence of fractional spot (0.5,1), indicated by the red circle. (a) 66 eV. (b) 87 eV.

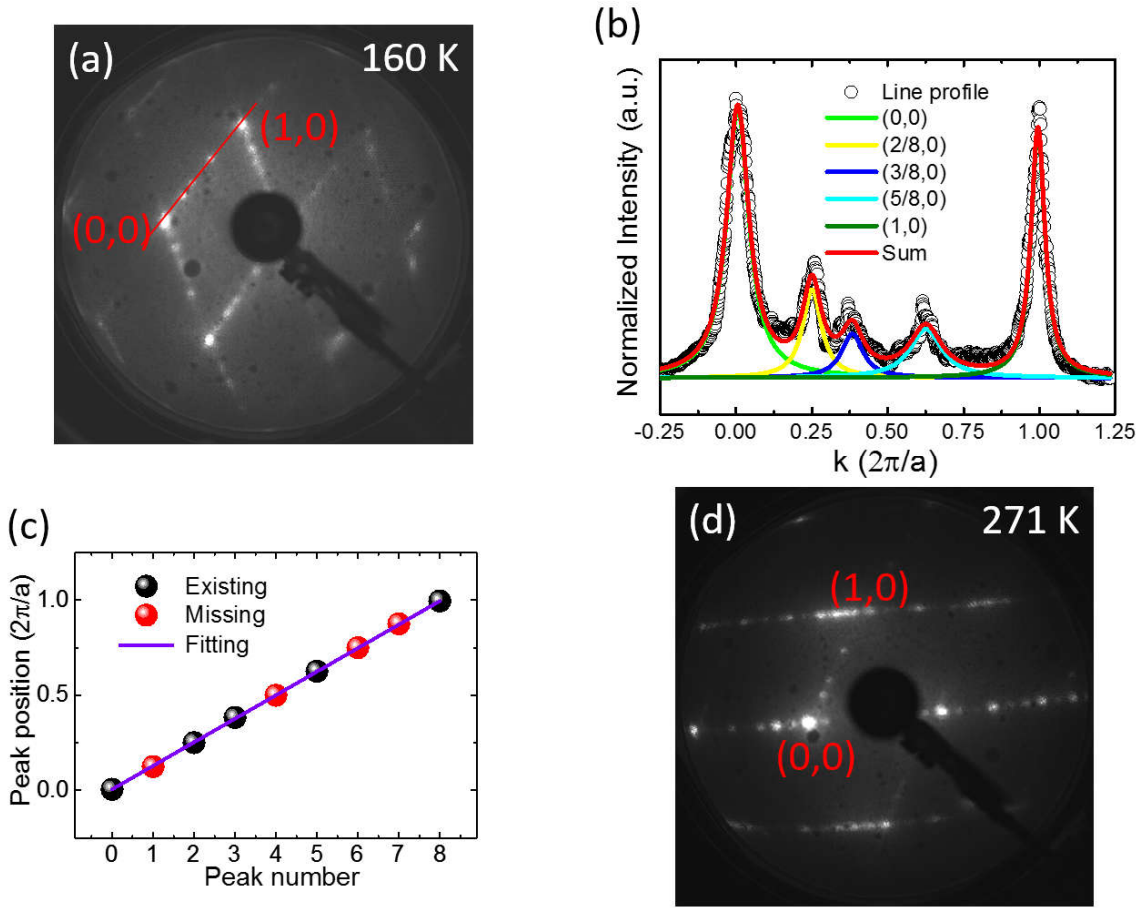

FIG. 8 (color online). Analysis of the 8×1 reconstructed pattern. (a) A reproduce of Fig. 3(a). (b) Line profile based on the red line in (a). The Lorentzian fittings of the peaks are also shown. (c) The peak positions of each spot with their peak numbers. The existing spots and missing spots are indicated by black and red colors respectively. The positions can be fit by a straight line. (d) LEED pattern at 271 K and 80 eV in the warming process. The sample is at the same position as Fig. 2(n) but rotated.